\documentclass[12pt]{article}
\usepackage{geometry} 
\usepackage{epsfig}               
\usepackage{graphicx}
\usepackage{amssymb}
\usepackage{cite}
\usepackage{amsmath}

\begin{document}

 \newcommand{\beq}{\begin{equation}}
\newcommand{\eeq}{\end{equation}}
\newcommand{\bea}{\begin{eqnarray}}
\newcommand{\eea}{\end{eqnarray}}
\newcommand{\beqn}{\begin{eqnarray}}
\newcommand{\eeqn}{\end{eqnarray}}
\newcommand{\beas}{\begin{eqnarray*}}
\newcommand{\eeas}{\end{eqnarray*}}
\newcommand{\defi}{\stackrel{\rm def}{=}}
\newcommand{\non}{\nonumber}
\newcommand{\bquo}{\begin{quote}}
\newcommand{\enqu}{\end{quote}}
\newcommand{\p}{\partial}


\def\de{\partial}
\def\Tr{ \hbox{\rm Tr}}
\def\const{\hbox {\rm const.}}
\def\o{\over}
\def\im{\hbox{\rm Im}}
\def\re{\hbox{\rm Re}}
\def\bra{\langle}\def\ket{\rangle}
\def\Arg{\hbox {\rm Arg}}
\def\Re{\hbox {\rm Re}}
\def\Im{\hbox {\rm Im}}
\def\diag{\hbox{\rm diag}}

\def\stroke{\vrule height8pt width0.4pt depth-0.1pt}
\def\topfleck{\vrule height8pt width0.5pt depth-5.9pt}
\def\botfleck{\vrule height2pt width0.5pt depth0.1pt}
\def\Zmath{\vcenter{\hbox{\numbers\rlap{\rlap{Z}\kern
0.8pt\topfleck}\kern
2.2pt\rlap Z\kern 6pt\botfleck\kern 1pt}}}
\def\Qmath{\vcenter{\hbox{\upright\rlap{\rlap{Q}\kern
3.8pt\stroke}\phantom{Q}}}}
\def\Nmath{\vcenter{\hbox{\upright\rlap{I}\kern 1.7pt N}}}
\def\Cmath{\vcenter{\hbox{\upright\rlap{\rlap{C}\kern
3.8pt\stroke}\phantom{C}}}}
\def\Rmath{\vcenter{\hbox{\upright\rlap{I}\kern 1.7pt R}}}
\def\Z{\ifmmode\Zmath\else$\Zmath$\fi}
\def\Q{\ifmmode\Qmath\else$\Qmath$\fi}
\def\N{\ifmmode\Nmath\else$\Nmath$\fi}
\def\C{\ifmmode\Cmath\else$\Cmath$\fi}
\def\R{\ifmmode\Rmath\else$\Rmath$\fi}


\def\QATOPD#1#2#3#4{{#3 \atopwithdelims#1#2 #4}}
\def\stackunder#1#2{\mathrel{\mathop{#2}\limits_{#1}}}
\def\stackreb#1#2{\mathrel{\mathop{#2}\limits_{#1}}}
\def\Tr{{\rm Tr}}
\def\res{{\rm res}}
\def\Bf#1{\mbox{\boldmath $#1$}}
\def\balpha{{\Bf\alpha}}
\def\bbeta{{\Bf\beta}}
\def\bgamma{{\Bf\gamma}}
\def\bnu{{\Bf\nu}}
\def\bmu{{\Bf\mu}}
\def\bphi{{\Bf\phi}}
\def\bPhi{{\Bf\Phi}}
\def\bomega{{\Bf\omega}}
\def\blambda{{\Bf\lambda}}
\def\brho{{\Bf\rho}}
\def\bsigma{{\bfit\sigma}}
\def\bxi{{\Bf\xi}}
\def\bbeta{{\Bf\eta}}
\def\d{\partial}
\def\der#1#2{\frac{\d{#1}}{\d{#2}}}
\def\Im{{\rm Im}}
\def\Re{{\rm Re}}
\def\rank{{\rm rank}}
\def\diag{{\rm diag}}
\def\2{{1\over 2}}
\def\ntwo{${\cal N}=2\;$}
\def\4N{${\cal N}=4$}
\def\none{${\cal N}=1\;$}
\def\x{\stackrel{\otimes}{,}}
\def\ba{\beq\new\begin{array}{c}}
\def\ea{\end{array}\eeq}
\def\be{\ba}
\def\ee{\ea}
\def\stackreb#1#2{\mathrel{\mathop{#2}\limits_{#1}}}

\def\Tr{{\rm Tr}}
\newcommand{\vp}{\varphi}
\newcommand{\pt}{\partial}

\setcounter{footnote}0

\vfill

\begin{titlepage}

\begin{flushright}
FTPI-MINN-06/02, UMN-TH-2427/06\\
ITEP-TH-01/06 \\
January 16, 2006
\end{flushright}

\begin{center}

{ \Large \bf Non-Abelian Strings and Axions}
\end{center}

\begin{center}
 { \bf A.~Gorsky\,$^{a,b}$,
 \bf    M.~Shifman$^{b}$ and \bf A.~Yung$^{a,b,c}$}
\end {center}
\vspace{0.3cm}
\begin{center}
$^a${\it Institute of Theoretical and Experimental Physics, Moscow
117259, Russia}\\
$^b${\it  William I. Fine Theoretical Physics Institute,
University of Minnesota,
Minneapolis, MN 55455, USA}\\
$^{c}${\it Petersburg Nuclear Physics Institute, Gatchina, St. Petersburg 188300, Russia}
\end{center}

\vspace*{.45cm}
\begin{center}
{\large\bf Abstract}
\end{center}
We address two distinct but related issues: (i)
the impact of (two-dimensional) axions in a two-dimensional theory
known to model confinement, the $CP(N-1)$ model; (ii) bulk axions 
in four-dimensional Yang--Mills theory
supporting non-Abelian strings.
In the first case $n,\,\,\bar n$ kinks play the role of ``quarks."
They are known to be confined. We show that introduction of
 axions leads to deconfinement
(at very large distances). This is akin to the phenomenon of
wall liberation in four-dimensional Yang--Mills theory.
In the second case we demonstrate that the bulk axion does {\em not} liberate
confined (anti)monopoles, in contradistinction with
the two-dimensional model. A novel physical effect which we observe is the axion
radiation caused by monopole-antimonopole pairs attached to the non-Abelian strings.

\vspace*{.05cm}

\end{titlepage}

\section{Introduction}
\label{one}

In this paper we address two distinct but related issues: (i)
the impact of (two-dimensional) axions in a two-dimensional theory
that is known to  model
confinement; (ii) axions in four-dimensional Yang--Mills theory
supporting non-Abelian strings which could play a role as cosmic strings.
In the first case we show that axions lead to deconfinement
(at very large distances). In the second case an interesting physical effect which we observe is the axion radiation caused by monopole-antimonopole pairs attached to the
non-Abelian strings.

As well known, two-dimensional $CP(N-1)$ model is an excellent
theoretical laboratory for modeling,
in a simplified environment, a variety of crucial phenomena
typical of  non-Abelian gauge theories in four dimensions, such as
confinement or chiral symmetry breaking
\cite{W79,NSVZsigma}.
Recently, two-dimensional $CP(N-1)$ model was shown to emerge
\cite {GSY05} as a moduli theory on the world-sheet of non-Abelian
flux tubes presenting solitons in certain four-dimensional Yang--Mills theories
at weak coupling \cite{HT1,ABEKY,SYcm,HT2}. This explains, in part,
a close parallel existing between  non-Abelian gauge theories in four dimensions
and two-dimensional $CP(N-1)$ models.

In this paper we will study an aspect of this parallel which so far escaped attention. Namely, we incorporate axions. Of course, everybody knows that
axions solve the strong $CP$ problem. This issue is not the focus
of our investigation, however. Our task is to study a less familiar phenomenon.

Let us start with pure Yang--Mills theory with the gauge group
SU$(N)$ (in what follows $N$ is supposed to be large, unless stated to the contrary).
As was shown by Witten \cite{W98}, in this theory there are
$\sim N$ quasi-stable vacua --- let us call them {\sl quasivacua} ---
entangled in the $\theta$ evolution. For each given $\theta$
one of states from this family is the true vacuum.
The energy densities of other quasivacua lie higher than that
of the true vacuum by $O(N^0)$. At the same time, the energy densities
themselves, as well as the barriers separating quasivacua, scale as $O(N^2)$.
The decay rate of the quasivacua
is exponentially suppressed at large $N$, see \cite{S99}.

\begin{figure}[h]
 \centerline{\includegraphics[width=3in]{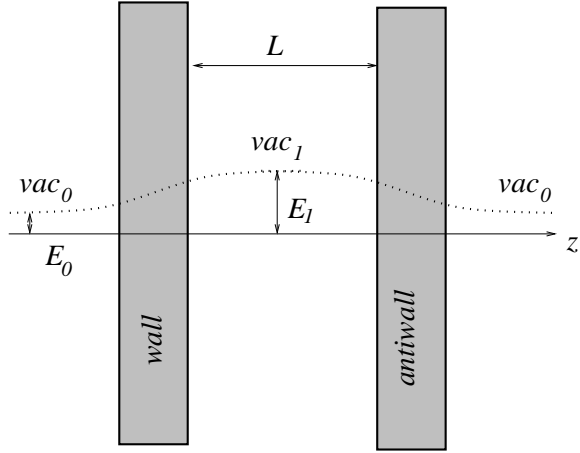}}
 \caption{\small The wall-antiwall configuration in pure Yang--Mills theory at large $N$.
 The notation is as follows:
 $vac_0$ stands for the lowest-energy state, while $vac_1$ is the adjacent quasivacuum. The corresponding energy densities are $E_0$ and $E_1$, respectively.
 The wall-antiwall pair experiences linear confinement with
  the energy of the wall-antiwall pair configuration growing as $(E_1-E_0)L$.
  }
 \label{aw}
 \end{figure}

Correspondingly,
one can expect occurrence of domain walls interpolating between the above vacua.
Because the latter are not exactly degenerate, strictly speaking, an isolated wall does not exist; rather, one must consider a wall-antiwall configuration (Fig.~\ref{aw}).
However, while the wall tension grows with $N$, the force between them
is $N$ independent.
Since this force is also independent of the distance between
the walls, one can speak of the wall-antiwall linear confinement,
albeit, this is a very weak $1/N$-suppressed confinement.

Now, this picture drastically changes once the axion field is added.
In Ref. \cite{Gab2000} (see also \cite{Gab2002}) it was shown
that the domain walls of  pure Yang--Mills theory
develop an axion component, axion tails with thickness $\sim m_a^{-1}$
where $ m_a$ is the axion mass.

What is most important,
the presence of the axion tails equalizes the vacuum energies
on both sides of the wall.
This is due to the fact that
one can view the axion field on the wall as an interpolation between
$\theta =0$ and $\theta = 2\pi$. As $\theta_{\rm eff}$
adiabatically changes, the vacua ``breathe" and effectively interchange
their energies in the course of interpolation. $E_1$ becomes equal to $E_0$.

This means that the walls become absolutely stable, even at
finite and not necessarily large $N$.
Each wall can be considered in isolation.  If we still consider the wall-antiwall
configuration with separation $\gg m_a^{-1}$, as in Fig.~\ref{aw},
there is no force between the wall and antiwall. In other words,
the wall confinement is gone at distances $\gg m_a^{-1}$.

In the $CP(N-1)$ model the role of walls is assumed by kinks.
 It was noted long ago \cite{W79} that at large $N$ the  $CP(N-1)$ model
is solvable within the framework of $1/N$ expansion, and this solution exhibits the following features:
(i) a kink mass term of order $\Lambda$ develops which does not scale with $N$
(here  $\Lambda$ is a dynamical scale parameter);
(ii) isolated kinks do not exist in the physical spectrum; the physical spectrum
is saturated by kink-antikink bound states;
(iii) there is a linear potential acting between kinks and antikinks; the slope of this potential is small at large $N$ since it scales as $\Lambda^2/N$.

A close analogy with the domain wall confinement in four dimensions 
is evident.\footnote{It is worth noting that lattice studies 
aimed at this question were reported in the literature recently \cite{lat}.}
The root of this analogy is similarity of the vacuum structure.
Much in the same way as in  pure Yang--Mills theory in four dimension, the
two-dimensional $CP(N-1)$ model
has $\sim N$ quasivacua which become stable and degenerate at
$N=\infty$. These vacua are all entangled in the $\theta$ evolution.

Below we will show that this analogy extends even further.
Namely, if the axion field is introduced in the $CP(N-1)$ model,
the kink-antikink confinement is eliminated. The force between
them vanishes for separations
$\gg m_a^{-1}$. The exact solvability of the
$CP(N-1)$ model allows us to describe this phenomenon
in fully quantitative terms using the framework developed in \cite{W79}.

In the second part of this paper we return to the four-dimensional bulk theory
supporting non-Abelian strings. In \cite{GSY05} it was shown
that the $\theta$ term of this theory
penetrates in the $CP(N-1)$ model on the world-sheet
of the non-Abelian string. Recently non-Abelian strings were suggested as candidates
for cosmic strings \cite{hashimoto}. Then it is natural to
promote the bulk $\theta$ term to a four-dimensional axion field
and discuss its impact on the non-Abelian strings. In particular, we will be mostly interested
in the fate of the kink-antikink pairs in the presence of the four-dimensional axion.
In fact, the kinks can be identified with
confined monopoles residing on the non-Abelian flux tubes \cite{tong,SYcm}.
It turns out that the four-dimensional axion, unlike its two-dimensional counterpart,
does not affect
confinement of the monopole-antimonopole pairs on the cosmic string.
The main effect due to four-dimensional axions is
that excitation of these pairs results in the axion radiation.
We will briefly comment on the emission of the ``cosmic" axions off
the non-Abelian strings.

The paper is organized as follows. In Sect.~\ref{aidc} we
consider two-dimensional axion in  the two-dimensional
$CP(N-1)$  model and show that the
axion-induced deconfinement of kinks takes place.
Section \ref{tri} is devoted to four-dimensional axion
interaction with non-Abelian flux tubes.
Orientational moduli of such strings are described
by the $CP(N-1)$ model on the string world-sheet.
After a brief outline of the basic bulk model (Sect.~\ref{bmna})
we introduce a bulk axion and
argue that the four-dimensional axion
does not liberate monopole-antimonopole pairs
attached to the string and confined inside ``mesons" (Sect.~\ref{mam}).
Section \ref{cnas}
treats the axion radiation
in the bulk in the context of non-Abelian cosmic strings.

\section{Axion induced deconfinement of kinks in two dimensions}
\label{aidc}

For our purposes the
most convenient formulation
of the $CP(N-1)$  model is in terms of the $n$
fields.\footnote{They are referred to as ``quarks" or solitons in
Ref.~\cite{W79}.}
 Then the $CP(N-1)$ model can be written as
\beq
{\cal L} = \frac{2}{g^2}\, \left[
\left(\partial_{\alpha} + i A_\alpha\right) n^*_{\ell}
\left(\partial_\alpha - i A_\alpha \right) n^{\ell}
-\lambda \left( n^*_{\ell} n^{\ell}-1\right)
\right]\,,
\label{onep}
\eeq
where $n^\ell$ is an $N$-component complex filed, $\ell = 1,2,...,N$, subject to the constraint
 \beq
n_{\ell}^*\, n^{\ell} =1\,.
\label{lambdaco}
\eeq
This constraint is implemented by the Lagrange multiplier $\lambda$
in Eq.~(\ref{onep}). The field $A_\alpha$ in this Lagrangian is also auxiliary,
it enters with no derivatives and can be eliminated
by virtue of the equations of motion,
\beq
A_\alpha =-\frac{i}{2}\, n^*_\ell \stackrel{\leftrightarrow}{\partial_\alpha} n^\ell\,.
\label{two}
\eeq
Substituting Eq.~(\ref{two}) in the Lagrangian, we rewrite it in the form
\beq
{\cal L} = \frac{2}{g^2}\, \left[
\partial_{\alpha}  n^*_{\ell}\,
\partial_\alpha  n^{\ell} + (n_\ell^*\partial_\alpha n^\ell)^2
-\lambda \left( n^*_{\ell} n^{\ell}-1\right)
\right]\,.
\label{three}
\eeq

Now, $g^2$ is the coupling constant. The factor of $2$ in the definition of
the coupling constant (see Eq.~(\ref{onep})) is introduced to
match the standard definition of the coupling constant in  the O(3) sigma model.
The coupling constant $g^2$ is asymptotically free, and defines
a dynamical scale of the theory $\Lambda$ through
\beq
\Lambda^2 = M_{\rm uv}^2 \exp\left(-\frac{8\pi}{Ng^2_0}\right)\,,
\label{seven}
\eeq
where $M_{\rm uv}$ is the ultraviolet cut-off and $g^2_0$
is the bare coupling.

At first, let us forget for
a while about the axion terms and outline
the solution of the ``axionless"  $CP(N-1)$ model
at large $N$ \cite{W79}.
To the leading order it is determined by one loop and can be summarized as follows:
the constraint (\ref{lambdaco}) is dynamically eliminated so that all $N$ fields
$n^\ell$ become independent degrees of freedom with the mass term $\Lambda$.
The photon field $A_\mu$ acquires a kinetic  term
\beq
{\cal L}_{\gamma\,\,\rm kin}= -\frac{1}{4e^2} F_{\mu\nu}^2\,,\qquad e^2 = \frac{12\pi \Lambda^2}{N}\,,
\label{pki}
\eeq
and also becomes ``dynamical." We use quotation marks here
because in two dimen\-sions the kinetic term (\ref{pki})
does not propagate any physical degrees of freedom; its effect
reduces to an instantaneous
Coulomb interaction. This is best seen in the $A_1=0$ gauge.
In this gauge the above kinetic term takes the form
\beq
(\partial_z A_0)^2
\eeq
while the interaction is
\beq
A_\alpha J^\alpha =A_0 J^0\,,\qquad J^\alpha=  n^*_\ell \stackrel{\leftrightarrow}{\partial_\alpha} n^\ell\,.
\eeq
Since $A_0$ enters in the Lagrangian without time derivative,
it can be eliminated by virtue of the equation of motion leading
to the instantaneous Coulomb interaction
\beq
J_0 \, \partial_z^{-2}\,  J_0\,.
\label{dev}
\eeq
In two dimensions the Coulomb interaction is proportional to
\beq
\frac{\Lambda^2}{N}\,  |z|\,.
\eeq
We get linear confinement acting between the $n,\,\,\bar n$ ``quarks."

The axion part of the Lagrangian can be written as follows:
\beq
{\cal L}_a = f_a^2\, (\partial_\mu a)^2 +\frac{a}{2\pi }\, \varepsilon_{\alpha\gamma}
\partial^\alpha A^\gamma\,,
\label{eight}
\eeq
where $A^\gamma$ is defined in Eq.~(\ref{two}), and $f_a$ is the axion constant.
In two dimensions it is dimensionless. As usual, the axion mass will be proportional to
$\Lambda/f_a$. We will consistently assume that $f_a\gg1$.

Upon field rescaling bringing kinetic terms to canonical normalization
one obtains
\beq
- \frac{1}{4} F^2_{\mu\nu} + \frac{e}{2\pi f_a}\, a\, \varepsilon_{\alpha\gamma}
\partial^\alpha A^\gamma +
(\pt_\mu a)^2+ e\, A_\alpha J^\alpha\,.
\label{dven}
\eeq
The axion field represents a single degree of freedom. The role of the ``photon" is that
upon diagonalization we get a massive spin-zero particle,
with mass of order $f_a^{-1}\Lambda N^{-1/2}$.

Since the exchanged quanta are massive
the long distance force responsible for confinement disappears,
giving place to deconfinement at distances $\gg m_a^{-1}$.
\begin{figure}[h]
 \centerline{\includegraphics[width=3in]{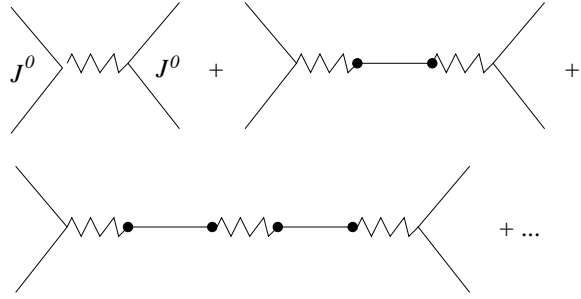}}
 \caption{\small The photon-axion mixing. Wavy lines denote photons, solid thin lines axions. Closed circles stand for the photon axion coupling
 $e/(2\pi f_a)$ where $e$ is defined in Eq.~(\ref{pki}).  The photon
 are coupled to to the currents (thick solid lines)  with the coupling constant $e$.}
 \label{gsy1}
 \end{figure}
Taking account of the photon-axion mixing amounts to summing the infinite series of
graphs depicted in Fig.~\ref{gsy1}. Using Eqs.~(\ref{dev}) and (\ref{dven})
it is not difficult to get for this sum
\beq
e^2J^0J^0 \left\{
\frac{1}{p^2} +\frac{1}{p^2}\left(\frac{e}{2\pi f_a}\right)^2\,\frac{1}{p_\mu^2} + ...\right\}=\frac{e^2J^0J^0}{p^2}\,\,\frac{p_\mu^2}{p_\mu^2 -\left(\frac{e}{2\pi f_a}\right)^2}\,.
\eeq
where $p$ is the spatial component of the momentum transfer $p_\mu$.
Using the current conservation one can rewrite
\beq
J^0J^0 \, p_\mu^2 = - p^2\, J_\alpha J^\alpha\,,
\eeq
which leads to the following final result for the sum
of the graphs depicted in Fig.~\ref{gsy1}:
\beq
- e^2J_\alpha J^\alpha\,\, \frac{1}{p_\mu^2 -\left(\frac{e}{2\pi f_a}\right)^2}\,.
\label{mpole}
\eeq
This expression is Lorentz invariant; it describes propagation of a quantum
of mass $e/(2\pi f_a )$. At distances larger than $2\pi f_a /e$
the force acting between $n$ and $\bar n$  is exponentially screened.
If instead of the emitter of ``photons" we consider an emitter of axions
and sum up the series of diagrams similar to that in Fig.~\ref{gsy1}
we arrive at the same pole as in Eq.~(\ref{mpole}). Of course, this is fully 
consistent with the fact that
the photon-axion system
in the problem at hand presents only one physical degree of freedom.

Note that the situation is very similar
to the {\em supersymmetric}
$CP(N-1)$ model whose physical spectrum  consists
of the kinks in the fundamental representation that
are not confined \cite{W79}. The reason for the kink liberation
in this case is as follows. The supersymmetric model involves
massless fermions which interact with the photon field
as $\bar{\psi}\gamma_{\mu} A_{\mu}\psi$.
The massless fermions can
be bosonized into the massless $\phi$ scalar with the
interaction term $\epsilon^{\mu\nu}A_{\mu}\partial_{\nu}\phi$,
which is the same as the photon-axion interaction in Eq.~(\ref{dven}).

The axion-induced liberation of kinks at distances $\gg m_a^{-1}$ we have just demonstrated is the two-dimensional counterpart
of domain-wall deconfinement in four-dimensions \cite{Gab2000,Gab2002}.
The parallel becomes even more pronounced in the (string-inspired)
formalism which ascends to \cite{susskind} (in connection with 
walls it was developed in \cite{Gab2000} and 
recently discussed in \cite{dvali} in another context). In this  formalism one introduces an
(auxiliary) antisymmetric three-form
gauge field $C_{\alpha\beta\gamma}$, while the
four-dimensional axion is replaced by an
antisymmetric two-form field $B_{\mu\nu}$ (the Kalb--Ramond field).
In four dimensions the gauge three-form field has no propagating
degrees of freedom while the Kalb--Ramond field $B_{\mu\nu}$ presents
a single degree of freedom. 
The domain walls are the sources for $C_{\alpha\beta\gamma}$, much in the same way as the kinks are the sources for $A_0$ in two dimensions. The field strength four-form built from $C_{\alpha\beta\gamma}$
is constant (cf. $F_{01}$ in two dimensions).
The $CB$ mixing produces one massive physical degree of freedom,
four-dimensional massive axion. Simultaneously,
the domain-wall confinement is eliminated at distances  $\gg m_a^{-1}$.
In full analogy with two-dimensional $CP(N-1)$, supersymmetrization of
Yang--Mills theory leads to wall deconfinement without
axion's help.

Can one understand this phenomenon in the language we used in Sect.~\ref{one}
for description of the wall confinement/deconfinement?
The answer is yes, the underlying physics is basically the same.
In the ``axionless" $CP(N-1)$ model there are
 $\sim N$ quasivacua  split in energy,
the splitting being
of order of $\Lambda^2/N$ (labeled by an integer $k$). Only the lower minimum is the true vacuum while
all others are metastable exited states. [In the large $N$ limit
the  decay rate is exponentially small, $\sim \exp(-N)$.]
 At large $N$,
the $k$ dependence of the energy density on the quasivacua, as well as the $\theta$ dependence, is
well-known
\beq
{ E}_k (\theta) \sim \, N\, \Lambda^2
\left\{1 + {\rm const}\,\left(\frac{2\pi k +\theta}{N}
\right)^2
\right\}
\,.
\label{split}
\eeq
At $\theta = 0$ the genuine vacuum corresponds to $k=0$,
while the first excitation to $k=-1$. At $\theta =\pi$
these two vacua are degenerate, at $\theta = 2\pi$
their roles interchange.

The energy split ensures
 kink confinement: kinks do not exist as  asymptotic states --- instead, they form
kink-antikink mesons. The regions to the left of the kink and to the right of the antikink
are the domains of the true vacuum (at $\theta =0$ it corresponds to $k=0$.)
The region between the kink and antikink
is an insertion of the adjacent quasivacuum  with $k=-1$.

When we introduce the axion, the vacuum angle $\theta$ is replaced by
a dynamical field, $a(t,z)$.
In the regions to the left of the kink and to the right of the antikink
$\langle a\rangle =0$.
If the region between the kink and antikink is large enough, $L\gg m_a^{-1}$,
the axion field in this region adjusts itself in such a way as to minimize energy,
$$
\langle a\rangle =0 \longrightarrow \langle a\rangle = 2\pi\,.
$$
This axion tail equalizes the energy densities to the left and to the right of
the kink (as well as to the left and to the right of
the antikink) liberating $n$ and $\bar n$ from confinement.

\section{Four-dimensional axion and non-Abelian\\ strings}
\label{tri}

In Sect.~\ref{aidc} we showed that introducing two-dimensional
axion in (nonsupersymmetric) $CP(N-1)$ liberates kinks.
Now, let us address another aspect. Let us
 introduce a four-dimensional axion in the bulk theory which supports
non-Abelian strings and confined monopoles  seen as kinks
in the world-sheet theory (also $CP(N-1)$). Then we study the impact of this four-dimensional axion
on dynamics of strings/confined monopoles.

\subsection{The bulk model with non-Abelian strings}
\label{bmna}

First, let us briefly outline the model.
Following \cite{GSY05} we consider a nonsupersymmetric model which is in fact
a bosonic truncation of ${\cal N}=2$ model.
It supports
non-Abelian strings. The key requirement
is the existence of  color-flavor locking
which provides  topological stability to the
stringy solutions of first-order equations analogous to BPS equations of the supersymmetric ``parent."
The action  has the form \cite{GSY05} (in the Euclidean notation)
\beqn
S &=& \int {\rm d}^4x\left\{\frac1{4g_2^2}
\left(F^{a}_{\mu\nu}\right)^{2}
+ \frac1{4g_1^2}\left(F_{\mu\nu}\right)^{2}
 \right.
 \nonumber\\[3mm]
&+&
 {\rm Tr}\, (\nabla_\mu \Phi)^\dagger \,(\nabla^\mu \Phi )
+\frac{g^2_2}{2}\left[{\rm Tr}\,
\left(\Phi^\dagger T^a \Phi\right)\right]^2
 +
 \frac{g^2_1}{8}\left[ {\rm Tr}\,
\left( \Phi^\dagger \Phi \right)- N\xi \right]^2
 \nonumber\\[3mm]
 &+&\left.
 \frac{i\,\theta}{32\,\pi^2} \, F_{\mu\nu}^a \tilde{F}^{a\,\mu\nu}
 \right\}\,,
\label{redqed}
\eeqn
where $T^a$ stands for the generator of the gauge SU($N$),
\beq
\nabla_\mu \, \Phi \equiv  \left( \partial_\mu -\frac{i}{\sqrt{ 2N}}\; A_{\mu}
-i A^{a}_{\mu}\, T^a\right)\Phi\, ,
\label{dcde}
\eeq
and $\theta$ is the vacuum angle, to be promoted to the axion field,
\beq
\theta \to \theta +a \to a(x) \,.
\label{promo}
\eeq
 The last term forces $\Phi$ to develop a vacuum
expectation value (VEV) while the last but one term
forces the VEV to be diagonal,
\beq
\Phi_{\rm vac} = \sqrt\xi\,{\rm diag}\, \{1,1,...,1\}\,.
\label{diagphi}
\eeq
This VEV results in the spontaneous
breaking of both gauge and flavor SU($N$)'s.
A diagonal global SU($N$) survives, however,
namely
\beq
{\rm U}(N)_{\rm gauge}\times {\rm SU}(N)_{\rm flavor}
\to {\rm SU}(N)_{\rm diag}\,.
\eeq
Thus, color-flavor locking takes place in the vacuum.

One can
combine the $Z_N$ center of SU($N$) with the elements
$\exp (2\pi i k/N)\in$U(1) to
get a topologically stable string solution \cite{HT1,ABEKY}
possessing both windings, in SU($N$) and U(1) since
\beq
\pi_1 \left({\rm SU}(N)\times {\rm U}(1)/ Z_N
\right)\neq 0\,.
\eeq
Their tension is $1/N$-th of that of the ANO string hence
the ANO string can be considered as a bound state of
$N$ elementary strings. These elementary strings are  $Z_N$ strings.

The most important feature of $Z_N$ strings in the model
(\ref{redqed}) is that they acquire orientational zero modes
associated with rotation of their color flux inside the non-Abelian
subgroup SU($N$) of the gauge group \cite{HT1,ABEKY,SYcm,HT2}. This makes
these strings genuinely non-Abelian.
This means that the effective low-energy theory on the string world-sheet
includes both the standard  Nambu-Goto action associated with translational
moduli and a sigma model which describes internal dynamics of the orientational
moduli.

As was mentioned above,
the emerging world-sheet action can be identified as the
$CP(N-1)$ sigma model \cite{HT1,ABEKY,SYcm,HT2}
 which (in the Euclidean notation)
is given by the action
\beq
S^{(1+1)}=   \int d t\, dz \,\left\{2 \beta\,
\left[(\pt^k\, n^*\pt_k\, n) + (n^*\pt_k\, n)^2\right]-
\frac{\theta +a}{2\pi}\,\,\varepsilon^{nk}\,\pt_n\, n^*\pt_k\, n
\right\}\,,
\label{cpN}
\eeq
where $\theta$ coincides with the four-dimensional $\theta$
while
\beq
\beta=\frac{2\pi}{g^2_2}.
\label{beta}
\eeq

The bulk theory is fully Higgsed, hence,    the monopoles are in the confinement
phase. In fact, as it was shown in \cite{tong,SYcm,GSY05}, they manifest themselves as junctions
of distinct elementary non-Abelian strings. In the string world-sheet theory
(\ref{cpN}) they are seen as kinks $n$ and $\bar n$
interpolating between the true vacuum and the adjacent quasivacuum of the
$CP(N-1)$ model (each quasivacuum in the $CP(N-1)$ model corresponds to a particular elementary string).

\begin{figure}
\epsfxsize=8cm
\centerline{\epsfbox{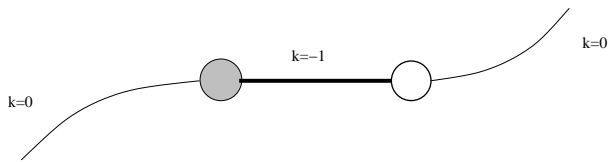}}
\caption{\small Monopole-antimonopole meson attached to the string. The filled
circle denotes the monopole while the empty circle  antimonopole.
The thick line  between the monopole and antimonopole denotes the
region of an exited string with $k=-1$.
}
\label{fig:meson}
\end{figure}

From the four-dimensional point of view this means that, besides
four-dimensional confinement, the monopoles are confined also in the two-dimensional
sense: if a monopole  is attached to a string, with necessity there is an antimonopole
 attached to the same string, and they
 form a meson-like configuration on the string
 \cite{MMY,GSY05}, see Fig. \ref{fig:meson}.

\subsection{Monopole-antimonopole ``mesons" vs. axion clouds}
\label{mam}

In this section we address the question what happens with the
monopole-antimono\-pole meson on the non-Abelian string in the presence of
the four-dimensional axion.
A priori  one might suspect that the
four-dimensional axion induces deconfinement of  monopoles
localized on the non-Abelian string, much in the same way
as the two-dimensional axion. Below we show  that this does not happen.

The classical action of the four dimensional bulk axion field is
\beq
L_{a}=\int d^4 x \left[ f_{a}^2(\partial a)^2 + \frac{ia}{32\pi^2}
F^{a}_{\mu\nu}\tilde{F}^{a}_{\mu\nu}\right]\,,
\eeq
where in the case at hand $f_a$ has dimension of mass. The axion  has a
small mass generated by four-dimensional bulk instantons
\beq
m_a^2\sim \frac{\Lambda^4_4}{f_a^2}\,\left(\frac{\Lambda_4}{\sqrt{\xi}}
\right)^{b-4},
\label{axionmass}
\eeq
where $b$ is the first coefficient of the $\beta$ function in the
theory (\ref{redqed}). As usual, it is assumed that $f_a\gg \Lambda_4$.

The impact of the  bulk axion on non-Abelian string is two-fold.
First, the axion gets coupled  to the translational
moduli of the string. Assuming that
the string collective coordinates adiabatically depend on the world-sheet coordinates
we  get for this coupling
\beq
{\cal L}_a^{(1)}\sim \xi\int d^4 x\, a(x)\,
\varepsilon^{ij}\,\varepsilon_{\alpha\beta}\,
\partial_{i}x^{\alpha} \, \partial_{j}x^{\beta} \,\,
\delta^{(2)}(x-x_{\rm string}(t,z)),
\label{bterm}
\eeq
where the indices $i,j= 0,3$ run over the string world sheet coordinates
 while the indices
$\alpha,\beta =1,2$ are orthogonal to the string world-sheet. One could  rewrite this expression in the covariant form trading the axion field for the Kalb-Ramon two-index field
$B_{\mu\nu}(x)$. However, for our purposes this is not necessary. The coupling
(\ref{bterm}) is not specific for non-Abelian strings, it
 is generated in the case of the Abelian (or $Z_N$) strings as well.

 Now, let us discuss compact orientational moduli.
It is easy to see that   no mixed $n$-$x$ terms appear in the axion Lagrangian
(at least, in the  the quadratic order in derivatives).
The bulk axion  generates a quadratic in $n$ coupling, as is clearly
seen from Eq.~(\ref{cpN}).
The impact of this term in the axion Lagrangian can be summarized as follows:
\beq
{\cal L}_a^{(2)}\sim\int d^4x\,a(x)\,
\varepsilon^{nk}\,\,\pt_n\, n^*\pt_k\, n\,\, \delta^{(2)}(x-x_{\rm string}(t,z)),
\label{axionpot}
\eeq

Consider the monopole-antimonopole pair attached to the string, as in Fig.~\ref{fig:meson}, where the axion is switched off.
For $k=0$ the string is in the state with the lowest energy.
The monopole-antimonopole meson on the string corresponds to
the region of the excited $k=-1$ string between the   monopole
and antimonopole while to the left of the monopole and to the right
of the antimonopole $k=0$. The energy of this meson is of order of
$
(\Lambda^2/N)\,L,
$
where $L$ is the distance between the monopole and antimonopole along the string.

Now, let us switch on the four-dimensional axion field.
What could happen (but, in fact, does not happen)
is that the axion field could develop a non-vanishing expectation value
$a=2\pi$ on the string between the monopole and antimonopole positions,
equalizing the string energies and thus
screening the confinement force. This is exactly what happened for
the two-dimensional
axion studied in Sect.~\ref{aidc}.

\begin{figure}
\epsfxsize=8cm
\centerline{\epsfbox{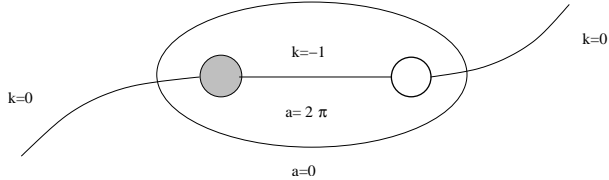}}
\caption{\small The
monopole-antimonopole meson together with the axion cloud. The region
of the string between the
monopole and antimonopole is not exited because although this region has $k=-1$
the  value of the axion field is non-zero inside the axion cloud, $a=2\pi$.
}
\label{fig:cloud}
\end{figure}

To see whether or not a similar effect occurs with four-dimensional
axion we have to examine a field configuration in which $\langle a\rangle =0$
everywhere in the bulk except a region adjacent to the
monopole-antimonopole separation interval, as depicted in Fig.~\ref{fig:cloud}.
We have to check the energy balance assuming there is an axion cloud
such that on the string inside the
monopole-antimonopole separation interval $\langle a\rangle =2\pi$,
which would let (anti)monopoles
attached to the string move freely along the string, with no confinement
along the string.

It is not difficult to estimate the energy of the axion cloud.
Transverse size of the cloud (in two directions perpendicular to
the string) must be of order of $m_a^{-1}$. The longitudinal dimension is $L$, see
Fig.~\ref{fig:cloud}.
Assume that $L\gg m_a^{-1}$. Then we get
\beq
E_{\rm cloud}\sim f_a^2 \,L,
\label{axioncloud}
\eeq
to be compared to the energy
$(\Lambda^2/N)\,L$ of the monopole-antimonopole meson.

Since $f_a$ is supposed to be very large compared to $\Lambda$
we see that the energy of the
axion cloud (\ref{axioncloud}) is much larger than the energy of
monopole-antimonopole meson. Developing a compensating axion cloud is energetically
disfavored.
Therefore we conclude that there is no monopole
deconfinement driven by four-dimensional axion.

Another way to arrive at the same conclusion involves  consideration
of the propagator of the four-dimensional axion between two
points on the string world-sheet, along the lines of Ref.~\cite{dgp}. Analysis
we  performed
in the momentum space shows that there is no macroscopic region
with the two-dimensional behavior on this propagator.

\subsection{Cosmic non-Abelian string and axion emission}
\label{cnas}

Recently  it was suggested \cite{hashimoto} to consider non-Abelian strings
as cosmic string candidates. Therefore it is worth discussing
possible signatures of such non-Abelian strings. Obviously,
they can be excited in
collisions. Both, translational and orientational modes can be excited.
In the latter case one can think of production of energetic monopole-antimonopole pairs
attached to the string and bound in mesons by the
confining potential along the string, as described in Sect.~\ref{mam}.
In Ref.~\cite{hashimoto} it was suggested
that isolated monopoles (kinks) can be created
on cosmic strings nonperturbatively. It is obvious that in our
non-supersymmetric case only pair creation is allowed.
Let us return to Fig.~\ref{fig:meson}.
 On the part of the string between the monopole and antimonopole
 (the kink and antikink) the state of the string is described by
the quasivacuum with $k=-1$. In this state\,\footnote{Henceforth
we will omit the $N$ factors since $N$ is not expected
to particularly large in the context of the cosmic strings.}
 \beq
 \langle \varepsilon^{nk}\,\,\pt_n\, n^*\pt_k\, n\rangle \sim \Lambda^2/N\,.
 \eeq
The topological charge density is localized
in the domain of the excited part of the string, and is approximately constant in this domain.
Therefore, as is clear from Eq.~(\ref{axionpot}), this interval,
whose length $L$ oscillates in accordance with the monopole-antimonopole
motion, will serve as a source term in the equation for the axion field.
 Assume that the energy of the
kink-antikink
pair $E\gg\Lambda$ so that they can be treated quasiclassically.
The distance $L$ between the kink and antikink will oscillate between
$-L_0$ and $L_0$ where $L_0\sim  E/\Lambda^2$ with the frequency
$\omega \sim \Lambda^2/ E$,
\beq
L(t) = L_0 \, e^{i\omega t}\,.
\eeq
Therefore, for a distant observer the monopole-antimonopole
meson is seen as a point-like source with the
interaction term
\beq
{\Lambda^2}\, \int d^4 x \,a(x)\,L(t)\,\delta^{3}(r-r_0),
\label{pointsource}
\eeq
where $r_0$ is a position of the meson on the string. The intensity of the
 axion radiation
from this point-like source can be estimated as
\beq
I_a \sim \omega^2\, \frac{\Lambda^4 L^2_0}{f_a^2}\,\frac1{r^2}
\sim \omega^2\, \frac{E^2}{f_a^2}\,\frac1{r^2} \,,
\label{mesonrad}
\eeq
where $r$ is the distance to the observer.

Of course the string produces axion radiation also due to coupling
 with translational modes, Eq.~(\ref{bterm}). This radiation is seen as coming from
a linear source, and can be estimated (per unit length)  as
\beq
I_a\sim \frac{\xi^{1/2}}{f_a^2}\,\frac{E^2}{\ell^2}\,\frac{1}{\rho}\,.
\label{trrad}
\eeq
Here $\rho$ is the  distance from the string to the observer in the plane
orthogonal to the string,
$E$ is the total excitation energy
and $\ell$ is the length of the excited part of the string. 
This radiation is not specific for non-Abelian
strings. Abelian ($Z_N$) strings produce this radiation as well.

Thus we see that  the non-Abelian string is seen by a distant observer as a
linear source of the axion radiation (\ref{trrad}), with additional point-like
sources of the axion radiation (\ref{mesonrad}) located on the linear source
at the positions of the monopole-antimonopole
mesons.

The rate of the axion radiation depends of $f_a$. The oscillating
kink-antikink pair will shake off energy until annihilation. The time duration of the
monopole-antimonopole meson de-excitation 
can be estimated as $T\sim E^2 f_a^2$.

\section{Conclusion}

The existence of the axion is almost unavoidable
in the framework of string theory.
In this paper we discussed how axions affect dynamics of strongly interacting objects.
The first part of the paper is devoted to a toy two-dimensional model,
nonsupersymmetric $CP(N-1)$, in which axions produce a dramatic effect:
they liberate kinks (confined in the absence of axion)
at distances $\gg m_a^{-1}$. Thus, we observed a novel phenomenon of
axion-induced deconfinement
of kinks.

This phenomenon is akin to the domain wall liberation
by axions in four-dimensi\-onal (nonsupersymmetric) Yang--Mills theory
\cite{Gab2000,Gab2002}.

Proceeding to four dimensions, we introduced a bulk axion in
the ``benchmark" nonsupersymmetric model \cite{GSY05} supporting
non-Abelian strings. Unlike its two-dimen\-sional counterpart,
the four-dimensional axion does not lead to monopole deconfinement.

Considering non-Abelian strings in the context of
cosmic strings we discussed axion emission due to excitations of such strings.
The excitations which produce axion radiation are of two types:
(i) excitations of the translational modes (the shape of the string),
and (ii) production of energetic pairs of confined
(anti)monopoles. The latter is specific to
 the non-Abelian strings.
We estimated the intensity of the axion radiation off the string and the time duration of the de-excitation process.

\section*{Acknowledgments}

We are grateful to G. Gabadadze for useful discussions.

The work
of A.G. was supported in part by grants CRDF RUP2-261-MO-04
and RFBR-04-011-00646.
A.G. thanks FTPI  at the University of Minnesota
where a part of this work was done   for the kind hospitality and
support.  He also thanks KITP at UCSB
for the hospitality during the program {\sl Mathematical Structures
in  String Theory} supported by grant NSF PHY99-07949.
The work of M.S. was
supported in part by DOE grant DE-FG02-94ER408.
The work of A.Y. was  supported in part
by  FTPI, University of Minnesota, and by Russian State Grant for
Scientific Schools RSGSS-11242003.2.

\end{document}